\documentclass[manuscript]{acmart}
\usepackage{booktabs}
\usepackage{tabularx}
\usepackage{hyperref}
\usepackage{subcaption}
\usepackage{multirow} 

\AtBeginDocument{%
  }

\setcopyright{acmlicensed}
\copyrightyear{2024}
\acmYear{2024}
\acmDOI{XXXXXXX.XXXXXXX}

\acmConference[Conference acronym 'XX]{Make sure to enter the correct
  conference title from your rights confirmation emai}{June 03--05,
  2018}{Woodstock, NY}
\acmISBN{978-1-4503-XXXX-X/18/06}

\newcommand{\hrvtablenew}{

\begin{table}[ht]
\resizebox{0.7\textwidth}{!}{%
\begin{tabular}{@{}lllll@{}}
\toprule
\multicolumn{1}{c}{\multirow{2}{*}{\textbf{HRV measures}}} &
  \multicolumn{1}{c}{\textbf{Card sequences}} &
  \multicolumn{1}{c}{\textbf{Button sequences}} &
  \multicolumn{1}{c}{\textbf{UI navigation}} &
  \multicolumn{1}{c}{\textbf{Slingshot}} \\ \cmidrule(l){2-5} 
  \multicolumn{1}{c}{\textit{}} & \multicolumn{4}{c}{\textit{Easy vs Challenging (t21)}}    \\ \hline
MeanRR  & -0.03 (0.97) & -0.23 (0.82) & -0.01 (0.99) & -0.40 (0.69) \\
SDNN    & 0.46 (0.65)  & 0.73 (0.47)  & 0.95 (0.35)  & 0.90 (0.37)  \\
RMSSD   & 0.06 (0.95)  & 0.29 (0.77)  & 0.55 (0.59)  & 0.77 (0.44)  \\
LF (nu) & 1.06 (0.30)   & -0.78 (0.44) & 0.55 (0.59)  & 0.52 (0.61)  \\
HF (nu) & -1.06 (0.30)  & 0.78 (0.44)  & -0.55 (0.59) & -0.52 (0.61) \\
LF/HF   & 1.06 (0.30)  & -1.00 (0.33)   & 0.83 (0.42)   & 0.81 (0.42)  \\
SD1     & 0.07 (0.95)  & 0.29 (0.77)    & 0.55 (0.59)  & 0.77 (0.45)  \\
SD2     & 0.72 (0.48)  & 0.94 (0.36)  & 1.25 (0.22)   & 0.94 (0.35)   \\ \bottomrule
\end{tabular}%
}
\caption{T-test results for different heart rate variability (HRV measures between the \textit{easy} and \textit{challenging} conditions. Values in brackets denote p-values.}
\label{tab:hrv}
\end{table}
}

\newcommand{\modeltab}{
\begin{table}[ht]
\resizebox{0.8\textwidth}{!}{%
\begin{tabular}{@{}lllll@{}}
\toprule
\multicolumn{1}{c}{\multirow{2}{*}{\textbf{}}} &
  \multicolumn{1}{c}{\textbf{Card sequences}} &
  \multicolumn{1}{c}{\textbf{Button sequences}} &
  \multicolumn{1}{c}{\textbf{UI navigation}} &
  \multicolumn{1}{c}{\textbf{Slingshot}} \\ \cmidrule(l){2-5} 
\multicolumn{1}{c}{} & \multicolumn{4}{c}{\textit{Within-user cross-validation accuracy (F1 score)}} \\ \midrule
Task condition       & .88 (.88)         & .85 (.86)         & .91 (.91)         & .87 (.87)         \\
Mental demand        & .71 (.83)         & .67 (.80)         & .71 (.08)         & .67 (.19)         \\
Temporal demand      & .63 (.77)         & .69 (.82)         & .55 (.66)         & 55 (.70)          \\
Frustration          & .51 (.55)         & .58 (.32)         & .60 (.22)         & .66 (.15)         \\
Arousal              & .56 (.70)         & .67 (.80)         & .71 (.83)         & .87 (.93)         \\
Valence              & .50 (.55)         & .65 (.79)         & .53 (.66)         & .86 (.93)         \\  \hline
\midrule
\multicolumn{1}{c}{} & \multicolumn{4}{c}{\textit{Leave-one-user-out cross-validation accuracy (F1 score)}} \\ \midrule
Task condition       & .45 (.52)         & .44 (.43)         & .43 (.30)         & .45 (.12)         \\
Mental demand        & .64 (.73)         & .67 (.64)         & .62 (.05)         & .62 (.05)         \\
Temporal demand      & 59 (.50)          & .71 (.76)         & .42 (.24)         & .50 (.34)         \\
Frustration          & .50 (.19)         & .51 (.00)         & .65 (.18)         & .65 (.21)         \\
Arousal              & .72 (.45)         & .77 (.67)         & .55 (.46)         & .88 (.90)         \\
Valence              & .51 (.41)         & .63 (.67)         & .60 (.55)         & .86 (.89)         \\
\hline
\end{tabular}%
}
\caption{Summary of Support Vector Classification (SVC) results.}
\label{tab:modelresults}
\end{table}
}

\begin{document}

\title{Motion as Emotion: Detecting Affect and Cognitive Load from Free-Hand Gestures in VR}
\renewcommand{\shorttitle}{Motion as Emotion}

\author{Phoebe Chua}
\affiliation{%
  \institution{Augmented Human Lab, National University of Singapore}
  \country{Singapore}}
\email{pchua@nus.edu.sg}

\author{Prasanth Sasikumar}
\affiliation{%
  \institution{Augmented Human Lab, National University of Singapore}
  \country{Singapore}}
\email{prasanth@ahlab.org}

\author{Yadeesha Weerasinghe}
\affiliation{%
  \institution{Augmented Human Lab, National University of Singapore}
  \country{Singapore}}
\email{yadeesha@ahlab.org}

\author{Suranga Nanayakkara}
\affiliation{%
  \institution{Augmented Human Lab, National University of Singapore}
  \country{Singapore}}
\email{scn@nus.edu.sg}

\renewcommand{\shortauthors}{Chua et al.}


\begin{abstract}
Affect and cognitive load influence many user behaviors. In this paper, we propose \textit{Motion as Emotion}, a novel method that utilizes fine differences in hand motion to recognise affect and cognitive load in virtual reality (VR). We conducted a study with 22 participants who used common free-hand gesture interactions to carry out tasks of varying difficulty in VR environments. We find that the affect and cognitive load induced by tasks are associated with significant differences in gesture features such as speed, distance and hand tension. Standard support vector classification (SVC) models could accurately predict two levels (low, high) of valence, arousal and cognitive load from these features. Our results demonstrate the potential of \textit{Motion as Emotion} as an accurate and reliable method of inferring user affect and cognitive load from free-hand gestures, without needing any additional wearable sensors or modifications to a standard VR headset. 



\end{abstract}



\received{20 February 2007}
\received[revised]{12 March 2009}
\received[accepted]{5 June 2009}

\begin{teaserfigure}
  \vspace{0.5cm}
  \includegraphics[width=\textwidth]{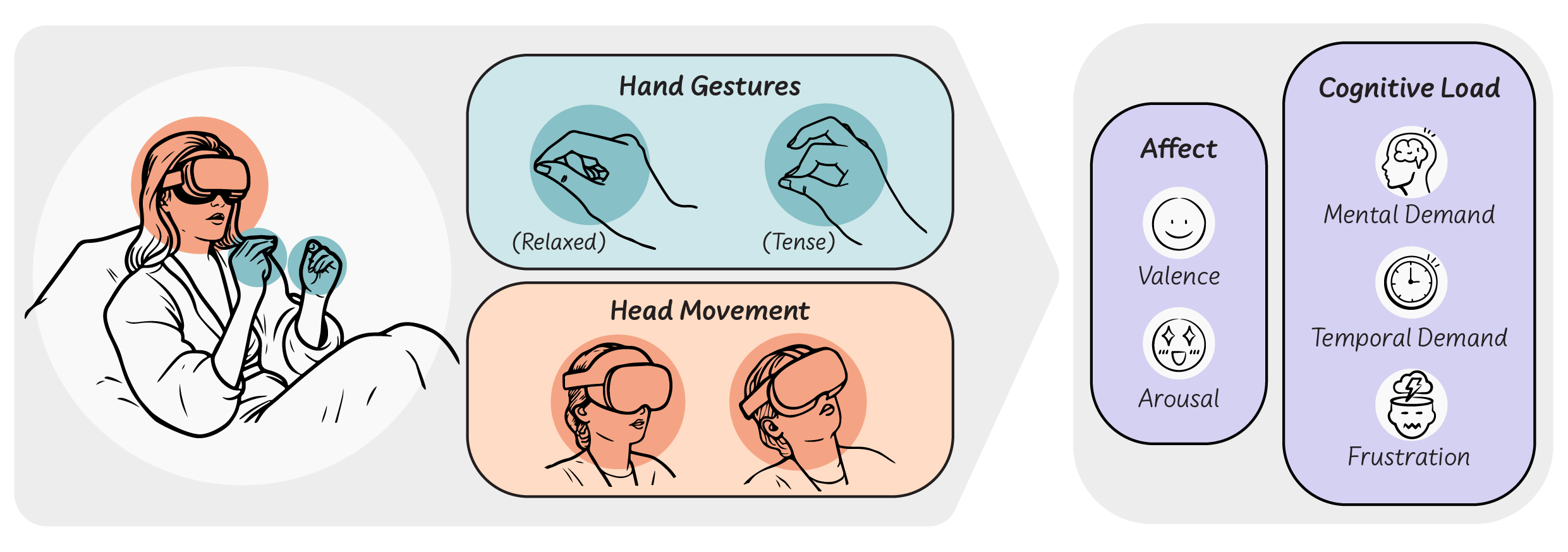}
  \caption{Our user study (n=22) investigates free-hand gesture inputs as a means of detecting user affect and cognitive load in VR. Across several VR tasks, we extracted features such as the speed and distance of participants' gesture formation, as well as their hand tension and head motion. From the features, we describe the relationship between affect, cognitive load and motion, and develop classification models to predict user affect and cognitive load.} 
  \label{fig:teaser}
  \Description{The figure depicts an individual wearing a VR headset and using free-hand gestures to interact with the virtual environment. Next to the individual, there are illustrations of hand and head movements that take place during these interactions. At the rightmost section of the figure, we include the terms valence, arousal, mental demand, temporal demand and frustration to provide an overview of the affective and cognitive states our study sets out to recognise from free-hand gestures.}
\end{teaserfigure}

\maketitle

\section{Introduction}

Systems that can identify and respond intelligently to a users' current emotional state and cognitive load benefit various domains, including online learning environments \cite{feidakis2016review}, gaming \cite{du2020non}, driving \cite{zepf2020driver} and e-commerce \cite{hibbeln2017your}. 
To this end, prior work has employed a variety of inputs for prediction ranging from facial expressions \cite{canal2022survey} and speech \cite{wani2021comprehensive} to physiological measures \cite{dzedzickis2020human} and notably, user-generated inputs from common human-computer interfaces such as keyboards and mice.

User input distinguishes itself from other forms of data traditionally used for emotion recognition by its availability, privacy-preserving nature and robustnesss. As a natural byproduct of human-computer interactions, it can be collected with no additional hardware, unlike facial expressions, speech and physiological data that often require external cameras, microphones and wearable sensors. Additionally, it is not reflective of personal identity in a way that might put an individual at risk if leaked or misused, giving it an advantage in privacy \cite{milne2017information, mcstay2020emotional}. Moreover, although physiological sensing is a widely used for recognition of cognitive and affective states in both research and consumer technologies, sensor data is known to be prone to artifacts and missing data \cite{foll2021flirt}. In comparison, the sensing principle behind user input relies on the finding that affect and cognitive load influence our muscle and attentional control. This allows detection of such states through keystrokes, mouse movements and other forms of user inputs, which are less susceptible to noisy recordings and may hence be able to capture more subtle changes in affect and cognitive load \cite{hibbeln2017your, goel2020stress, freeman2011hand, sun2014moustress}. 

The increasing popularity of immersive technologies for both consumer and enterprise applications presents an opportunity to investigate whether and how we can extend prior findings on emotion inference from user inputs from common human-computer interfaces to immersive environments. Heads-up computing devices such as VR headsets and smart glasses offer unique interaction techniques through free-hand interaction and specialized VR controllers \cite{luong2023controllers, zhao2023heads}, and in particular, free-hand interaction techniques that are driven by technologies for precise gesture and hand pose recognition generate novel types of input data that have yet to be explored in the context of user emotion recognition. Moreover, as an input modality, free-hand interactions have several advantages over VR controllers: they require no external equipment, elicit stronger feelings of ownership and realism \cite{lin2019effect}, and richer user experiences \cite{pei2022hand}.


%
%
%

In this work, we extend the literature on understanding users by demonstrating the potential of using free-hand interactions in VR for inferring user affect and cognitive load. We first conducted a study to collect hand and head-tracking data from individuals engaging in VR tasks of varying difficulty. Our study gathered over two and a half hours of data from 22 participants whose experience ranged from never having used VR at all, to those who use VR several times a week. We find that participants' reported valence, arousal and cognitive load are associated with significant differences in the speed and distance of gesturing. Further analyses revealed that participants display increased hand tension and reduced head movements when engaging in more challenging tasks. To the best of the authors' knowledge, this is one of the first empirical study that investigates the relationship between affect, cognitive load and free-hand gesture inputs in VR. We contribute to research by: 
\begin{itemize}
\item Presenting a detailed empirical investigation of the changes in gesture and motion features that occur in free-hand VR interactions as a result of changes in user affect and cognitive load.
\item Offering an explanation for the results, drawing on theories of grounded cognition, attention and human performance in HCI.
\item Providing insights for implementing effective, physiologically-motivated measures of user affect and cognitive load that can be computed from common free-hand VR gestures.
\end{itemize}
\section{Background \& Related Work}


\subsection{The Mind-Body Connection}
Viewed through the lens of grounded cognition \cite{barsalou2008grounded} and attentional control theory \cite{eysenck2007anxiety}, how we move is deeply reflective of the mind in motion \cite{freeman2011hand}. The brain is constantly running internal simulations of the body in the world -- for example, perceiving a cup activates internal simulations of a grasping motion, in anticipation of picking up the cup \cite{tucker1998relations}. These internal simulations use cues from our perception (e.g., vision), motor and introspective systems to facilitate efficient regulation of resources \cite{barrett2017theory}. However, affect and cognitive load have been shown to modulate our \textit{attention} to such cues through the mechanisms of attentional inhibition (preventing less relevant stimuli from capturing attention) and attentional shifting (allocating attention to more relevant stimuli). When experiencing negatively-valenced emotions, people's attention tends to shift from being goal-directed to being stimulus-driven as the brain's attentional inhibition and shifting functions are suppressed \cite{hibbeln2017your}. Similarly, both low and high levels of arousal have been linked to reduced attentional control \cite{da2020wandering}. From a physical perspective, the neurotransmitters released during heightened stress and cognitive load have also been found to increase movement variability and muscle tension \cite{goel2020stress}. Taken together, these results suggest that movement can be viewed as an embodiment of cognitive, affective and attentional processes, and that changes in these processes leave detectable motor traces.


In the field of HCI, information-theoretic analogies have played a key role in shaping the community's understanding of human movement. Fitts' Law \cite{fitts1954information} suggests that there is an inherent tradeoff between the distance, speed and accuracy of movements (i.e., fine motor control) due to the finite bandwidth of the the neurological pathways through which movement information is processed and transmitted. As such, as the need for fine motor control increases, the duration or distance of movement should increase correspondingly. Fitts' law has been validated across a range of 1D ((e.g., knobs and sliders \cite{mackenzie1992extending}), 2D (e.g., mouse movements \cite{grimes2015mind}) and 3D inputs (e.g., gestures \cite{burno2015applying}) as well as control interfaces such as touchscreens \cite{xie2023fitts}, demonstrating its generalizability. The Hick-Hyman Law \cite{hick1952rate,hyman1953stimulus}, which was built on observations that it takes longer to respond to a stimulus when it belongs to a large set as opposed to a small stimuli set, similarly suggests that increased cognitive load decreases the bandwidth available for fine motor control. These theories highlight important perceptual and psychomotor principles that have underpinned the design of human-computer interactions.




Our hands are one of the richest sources of body language information \cite{noroozi2018survey} as well as one of the primary means for interacting with computing devices. Despite strong evidence that user input data is an effective avenue for inferring affective states and cognitive load \cite{yang2021review}, there has been limited work exploring the use of free-hand gesture inputs for recognising affect and cognitive load. Existing work on inferring emotions from hands have primarily been situated in the expressive gesture literature, and have largely focused on understanding the symbolic meanings of natural gestures used to express emotion in everyday life, such as the "thumbs up" gesture \cite{koh2019developing, luo2024emotion}. It is unclear whether and how these findings map to the more structured gestures typically used to facilitate control and navigation of XR environments.

\subsection{Detecting Affect and Cognitive Workload from User Inputs}

Affect, which is predominantly characterized by the dimensions of hedonic valence (pleasure/displeasure) and arousal (activation/sleepy), forms the core of conscious experience \cite{duncan2007affect}. Cognitive workload describes the mental effort required to complete a task (“work”) under several constraints (“load”) \cite{kosch2023survey}. In HCI research, the concepts of affect and cognitive workload have been widely used to evaluate the usability and user experience aspects of interactive computing systems, as well as to support the design of adaptive user interfaces. Automatic methods to infer user affect and cognitive workload have often utilized wearable devices like smart watches, rings, armbands and chest straps to monitor biosignals such as skin temperature, electrodermal activity (EDA), breathing rate, heart rate and heart rate variability (HRV), which are influenced by activity in the central nervous system (CNS) and autonomic nervous system (ANS) \cite{saganowski2022emotion}. Although sensor quality has improved greatly in recent years, wearable devices still tend to be susceptible to motion artifacts and missing data \cite{foll2021flirt}. Moreover, their use requires individuals to purchase and maintain an additional device. 


To address these issues, research has explored the use of alternative data sources such as user inputs to capture affect and cognitive workload. The vast majority of our interactions with computing devices involve sending inputs through the keyboard and mouse, trackpads, and touchscreens. Although our awareness of frequently-performed actions such as typing or moving the mouse often recedes into the background, freeing up mental resources for our primary task \cite{heidegger2010being, rosenberger2009sudden}, these common actions provide a great deal of information about our underlying affective and cognitive states. For instance, people click and type with greater force under stress \cite{gao2012does, hernandez2014under, exposito2018affective, goel2020stress}. Additionally, negatively-valenced emotion has been found to reduce speed and increase the distance of mouse cursor movements, due to slower reaction times and reduced precision of movement \cite{hibbeln2017your}. With keyboards, keystroke features such as dwell time, latency and heat maps have been used to successfully recognise affect and stress \cite{wampfler2022affective, epp2011identifying, pepa2020stress}, while on mobile phones, linear acceleration and rate of rotation as well as stroke length, speed and trajectory have been employed to recognise affect \cite{wampfler2022affective, gao2012does}. Although these prior works suggest the potential of user inputs for inferring user emotion, they have primarily explored user inputs from common personal computing devices such as mobile phones, laptops and PCs.

\subsection{Research Context: User Inputs in VR} 
Extended reality (XR) devices such as virtual reality headsets have improved dramatically in terms of usability and size, turning them into an increasingly popular consumer computing device. As an interface, they offer novel, immersive experiences in body-compatible form factors such as headsets and smart glasses \cite{zhao2023heads}. Given the potential of XR devices to become much more commonplace over the next few years, the ability to easily assess user emotions during XR system use has implications for the design of adaptive applications that can respond intelligently to the user.


Two prominent interaction modalities in XR are controller-based inputs and free-hand inputs \cite{luong2023controllers}. Controllers are typically equipped with buttons, joysticks and triggers that provide intuitive ways for users to navigate and perform actions in virtual environments. With free-hand inputs, users primarily interact with the contents of virtual environments by using their bare hands to form gestures that are recognized and used as control commands. 
Although controllers have historically been associated with better reliability and accuracy \cite{navarro2019evaluating},
free-hand gesture inputs have the potential to create compelling interactions in immersive environments, as demonstrated by the release of the Apple Vision Pro in late 2023. These free-hand gestures often mimic well-known real-world actions, making them intuitive and easy to learn \cite{lu2014hand}. Furthermore, unlike handheld controllers,
gesture interactions allow users to effortlessly switch between tasks in the physical and virtual realms \cite{pei2022hand}, giving them a strong advantage in the context of daily use.

In this study, we extend previous work on inferring affect and cognitive load from user inputs to virtual reality environments by investigating the use of free-hand gesture inputs. Gesture inputs are distinct from inputs from common personal computing devices in terms of the raw data that can be logged, as well as the features that can be extracted from the logged data. In terms of raw data, existing user inputs typically take the form of x and y coordinates, pressure estimates and touchscreen or keyboard activity.
Compared to interactions that operate primarily on 2D planes, gesture interactions in 3D space have an additional degree of freedom (depth). XR devices such as the Meta Quest, HTC Vive and Hololens headsets are capable of hand tracking through built-in cameras that estimate the 3D position and orientation of hand keypoints. Free-hand gesture inputs also place different biomechanical demands on the user compared to common computing devices, which may influence the motion features that are relevant to detecting changes in affect and cognitive load \cite{goel2020stress}. 

To our knowledge, only a few studies have examined how affect and cognitive load influence the usage and formation of expressive gestures \cite{luo2024emotion, mcneill2019gesture}. However, these works did not test with the more structured hand gestures used to facilitate interactions with immersive environments. We address this gap through a study with participants who engaged in several VR tasks, meant to induce varying levels of valence, arousal and cognitive load. Through the use of 3D hand and head tracking data, we we investigate the changes in gesture and motion features that accompany changes in participant affect and cognitive load. The following section describes the procedures of our user study in further detail.

\section{Method} \label{sec3:method}


\subsection{Task Design}

\begin{figure}[ht]
    \centering
    \includegraphics[width=\linewidth]{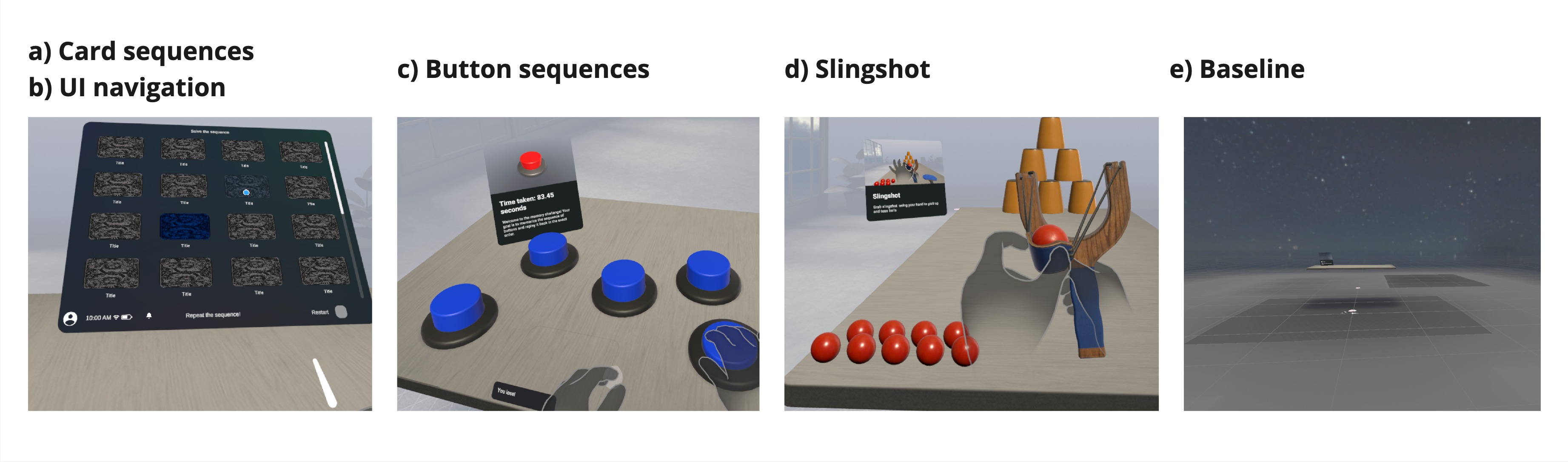}
    \caption{View of the virtual environment for each of the task conditions. Each task involves common interactions in VR, such as selection, swiping and realistic interactions with objects in the environment.}
    \label{fig:tasks}
    \Description{Interior scenes from the VR tasks used in this study. Scenes labeled "a" (card sequences) and "b" (UI navigation) depict a virtual menu with sixteen cards and a ray pointer extending from the users' hand. Scene labeled "c" (button sequences) depicts a virtual table with several buttons. Scene labeled "d" (slingshot) depicts a virtual slingshot and several balls on a table, with the user aiming at a tower of cups. Scene labeled "e" (baseline) depicts a starry night sky.}
\end{figure}

In deciding which interactions should be used in our study, we first considered existing VR use-cases. The most commonly cited reasons for using a VR device were gaming (72\%), watching films or TV (42\%), exercising (35\%) and browsing the internet (29\%)\footnote{https://www.nrgmr.com/our-thinking/technology/the-vr-revolution-might-finally-be-on-the-horizon/}. To enhance the generalizability and applicability of our results, we developed VR tasks that resembled common gaming and user interface navigation interactions, and designed conditions to induce varying levels of valence, arousal and cognitive load, which are frequently-studied mental states. 
We pretested several VR tasks in a pilot study to ensure that they elicited the expected affect and cognitive load. Four tasks and an additional baseline scene (Fig \ref{fig:tasks}) were selected for the final study. All tasks were designed in the Unity engine, and the duration of each task was standardized at one minute.

\textbf{Slingshot task:} The slingshot task was designed to elicit a sense of fun and enjoyment. To do so, it incorporates several elements that have been shown to make games enjoyable \cite{Schaffer2018WhatMG} including body movement, strategizing, clear goals and continuous feedback. The task environment consists of a long table, with a tower of six cups at one end and the user at the other. A slingshot and several balls are placed next to the user. Participants were instructed to knock down as many cups as possible using the slingshot. 
They received auditory feedback through sound effects when pulling the slingshot band, with the pitch increasing to signal increasing tautness. In the \textit{easy} condition, participants could aim for any cup in the tower. In the \textit{challenging} condition, participants were told that they would only gain points if they successfully knocked down the red cup in the tower. For each game, one cup was selected at random to be colored red.

\textbf{Card sequence memorization task:} The card sequence task was designed to induce high cognitive load and elicit stress. Memorization of visual patterns is a common approach used to increase cognitive load \cite{deck2021consistency}. The task interface consists of a 4x4 arrangement of 16 cards on a plane (see Fig. \ref{fig:tasks}). When the task begins, cards are highlighted in a sequence that is randomly generated without replacement. Each card in the sequence is highlighted for 0.5 seconds, with a 0.7-second interval between highlights. After the full sequence is presented, participants are tasked with selecting the cards in the same order in which they were highlighted. Selection could be performed in two ways: either by tapping their index and thumb together in a pinching gesture, which generated a virtual pointer, or by using their fingers to intersect the task plane. Participants received auditory feedback through sound effects if they successfully completed a sequence or if an incorrect card was selected. Following either of these events, a new sequence would begin after a 0.7-second delay. In the \textit{easy} condition, the sequence is of length 3; in the \textit{challenging} condition, the sequence is of length 6. The sequence length was initially chosen using average performance on the forward digit span task \cite{gregoire1997effect}, and adjusted based on results from the pilot studies such that the \textit{easy} condition clearly induces less cognitive load than the \textit{challenging} condition.

\textbf{Button sequence memorization task:} Like the card sequence task, the button sequence task was designed to induce high cognitive load and elicit stress. While the card sequence task primarily leverages the \textit{virtual pointer} metaphor, allowing participants to select cards using ray-pointing, the buttons were designed to leverage the \textit{virtual hand} metaphor \cite{luong2023controllers}, where the users interact with the environment through a virtual hand representation that mimics the movements of their physical hand. The task interface consists of several large buttons distributed randomly across a rectangular table. When the task begins, buttons are highlighted in a sequence that is randomly generated without replacement. After the full sequence is presented, participants are tasked with pressing the buttons in the same order in which they were highlighted. Button presses were triggered when participants use their fingers or palm to intersect with and press down on a button. The auditory feedback and sequence lengths used for the easy and challenging conditions were the same as that of the card sequence task.

\textbf{User interface (UI) navigation task:} The UI navigation task was designed to induce frustration. The UI navigation task features a six-page menu, with each page containing a 4x4 grid of 16 cards. The first and last pages of the menu always included one card highlighted in red. Participants were instructed to repeatedly swipe from the first to last page and back again, searching for and selecting the highlighted cards on each pass. A swipe was triggered by bringing the index and thumb together in a pinch, flicking the fingers to the left or right, and releasing the index and thumb in one smooth motion. We designed the easy and challenging conditions based on prior work exploring the reasons that lead to frustration \cite{bassano2019vr, roseman2004appraisals}. Specifically, events that are perceived as outside of one's control and inconsistent with personal motives are expected to elicit more frustration. In the \textit{easy} condition, vertical scrolling was locked. This had the effect of reducing participants' sense that they were "doing something wrong" (scrolling vertically when they were supposed to scroll across), and also of reducing visual distraction. In the \textit{challenging} condition, both vertical and horizontal scrolling was enabled. While we did not manipulate the task to artificially create vertical scrolling, the swiping gesture tended to be sensitive to vertical movement, making it more difficult to scroll across when vertical scrolling was not locked. 

\textbf{Baseline:} We designed a calming environment to measure participants' resting heart rate (HR) and heart rate variability (HRV). The environment features a view of a starry night sky set against soft background music with low complexity and acoustic variation. The music was chosen to help participants maintain a steady, regular breathing pattern \cite{avila2012influence}. This baseline allows us to better evaluate physiological changes that occurred when participants engaged in subsequent tasks \cite{shaffer2017overview}.


\subsection{Data Collection}

\subsubsection{Participants}
22 participants (11 females) aged between 19 and 45 years (mean = 29.2, SD = 7.30) were recruited from within the staff and student population of a university. The study was approved by the local ethics committee. All participants had no prior history of or tendency to experience motion sickness or vertigo. At the beginning of each study session, an experimenter introduced the participant to the study aims, the equipment involved and the data that would be recorded. Participants then gave informed consent and filled out a questionnaire to assess their demographic profile (age, gender, handedness), prior experience with VR and hand-tracking applications on a 5-point Likert scale (from 1 - Never used previously to 5 - Use a few times a week), familiarity with VR interfaces (1 - Novice, 2- Neutral, 3 - Expert).

\subsubsection{Procedure}
The study was conducted in a quiet, well-lit room in an academic department. The room was set up with a chair, table and laptop, which participants used to complete questionnaires. Once the study began, participants remained seated at a desk equipped with an RGB camera recording video at 1080p (1080 $\times$ 1920) at 30fps. Participants first removed all large accessories from their hands and wrists. Then, the experimenter would help participants to put on the Meta Quest Pro \footnote{https://www.meta.com/quest/quest-pro/} headset as well as the Polar Verity Sense. 

Participants were shown a trial scene where they could try each task as many times as they wanted. Once they felt comfortable with the gestures and tasks, we presented the baseline (calming) environment and asked participants to relax for 5 minutes while we recorded their resting heart rate and HRV. A 5-minute recording is the conventional standard for short-term HRV \cite{shaffer2017overview}.

We conducted the study using a within-subjects design in which participants performed all four tasks (slingshot, card sequence memorization, button sequence memorization and UI navigation) in both the \textit{easy} and \textit{challenging} conditions. The order of tasks and conditions was counterbalanced across participants.

\subsubsection{Measures}
We recorded three measures of mental state: subjective self-reports, physiological data, and hand tracking data. 

\textit{Subjective self-reports.}
After each task, participants reported their valence and arousal using the Affective Slider \cite{betella2016affective}, a validated self-reporting tool that adapts the popular Self-Assessment Manikin \cite{bradley1994measuring} for digital devices. The Affective Slider questions were administered on an 11-point rating scale (0-10). Subjective workload was assessed using the mental demand, temporal demand and frustration subscales of the NASA Task Load Index (TLX) \cite{hart1988development}. We also provided an optional open-ended question: "In your own words, how would you describe the task you just completed?"

\textit{Physiological data. }
To record participants' heart rate and heart rate variability (HRV), we used the Polar Verity Sense optical heart rate monitor, which has been shown to provide accurate continuous measurements of both heart rate and HRV \cite{topalidis2023pulses}. All participants wore the Polar Verity Sense on the bicep of their non-dominant arm to minimize motion artifacts. After removing outlier RR intervals, We extracted the HRV indicators in Table \ref{tab:hrv} using the hrvanalysis \footnote{\url{https://aura-healthcare.github.io/hrv-analysis/readme.html}} library.

\textit{Motion tracking data.}
To characterize users' motion, we logged the 3D positions and orientations of the head, 21 joints on the users' left and right hands, and their estimated pinch strength using the Meta Quest Pro's motion-tracking systems at a sampling rate of 20Hz. 







\section{Results}
We first report the 
affect and cognitive workload induced by each task. Then, we detail our findings from the quantitative analysis of freehand gesture inputs. Using the 3D hand tracking data collected, we train and evaluate linear models to perform binary classification of affective states and cognitive workload. Finally, we explore the possibilities for implicit human computer interaction \cite{schmidt2000implicit} during periods where a user is idle. All p-values were adjusted using the Benjamini-Hochberg procedure \cite{thissen2002quick} to control the false discovery rate.

\subsection{Affect and Cognitive Workload Elicited}


\begin{figure}[ht]
    \centering
    \includegraphics[width=\linewidth]{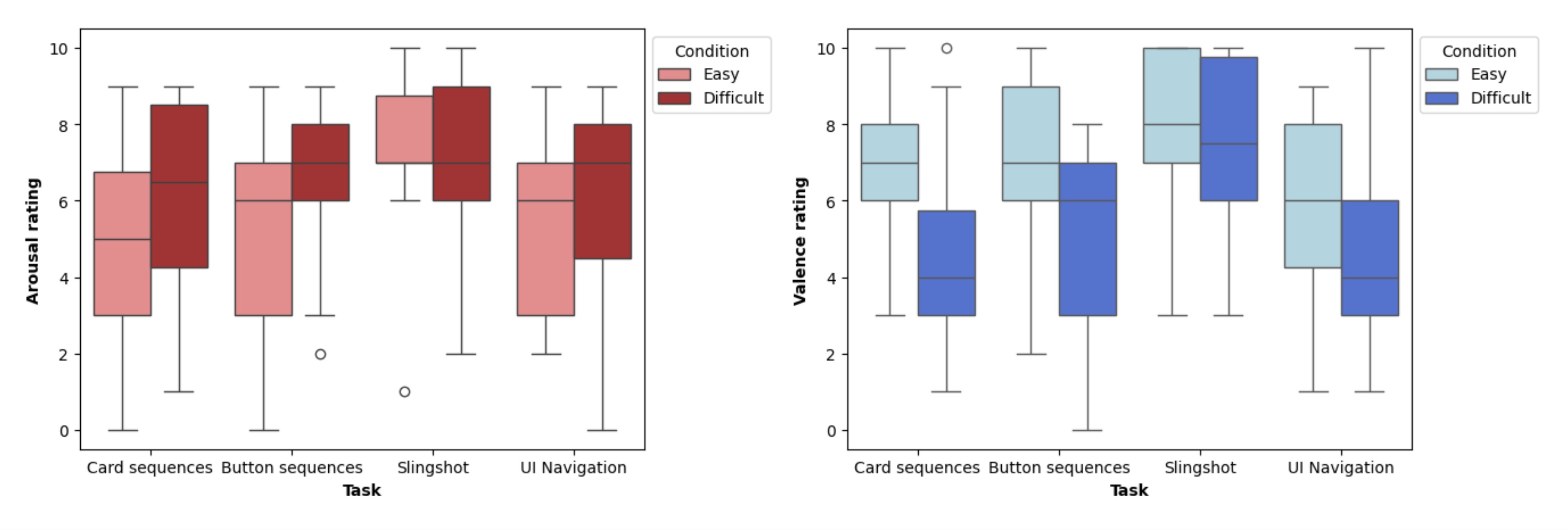}
    \caption{Effect of {\sc task condition} on induced arousal (left) and valence (right). Arousal is significantly higher in the \textit{challenging} condition for all except the Slingshot task; Valence is significantly lower in the \textit{challenging} condition for all except the Slingshot task.}
    \label{fig:task-aroval}
    \Description{This image consists of two side-by-side box plots. The left plot represents "Arousal rating" on the y-axis, and the right plot shows "Valence rating" on the y-axis. Both plots have "Task" on the x-axis, with the tasks labeled as "Card sequences," "Button sequences," "Slingshot," and "UI Navigation." In both plots, two conditions are represented: "Easy" (in lighter shades, pink on the left and light blue on the right) and "Difficult" (in darker shades, red on the left and dark blue on the right). In the arousal plot (left), higher arousal ratings appear for more difficult tasks compared to easier ones. In the valence plot (right), easy tasks generally show higher valence ratings (indicating more positive feelings), while difficult tasks show lower valence ratings. The boxes represent interquartile ranges, with the whiskers indicating the range of the data, and there are outliers shown as small circles.}
\end{figure}

\textit{Arousal and valence. } Here we describe the main results from the subjective self-reports obtained after each task to verify that the task conditions elicited the intended affective states. Where the assumptions about the normality of the data were violated (Shapiro-Wilk test \textit{p} $<$ .05), the data was transformed  using a log function prior to analysis. We conducted one-tailed paired t-tests for the button sequence, card sequence and UI navigation tasks and two-tailed paired t-tests for the slingshot task. Results showed that participants reported significantly higher arousal and lower valence in the challenging condition compared to the easy condition, for the button sequence (Arousal: \textit{t}(21) =  2.68, \textit{p} $<$ .05; Valence: \textit{t}(21) =  -5.29, \textit{p} $<$ .001), card sequence (Arousal: \textit{t}(21) =  2.03, \textit{p} $<$ .05; Valence: \textit{t}(15) =  -3.29, \textit{p} $<$ .01) and UI navigation (Arousal: \textit{t}(21) =  2.37, \textit{p} $<$ .05; Valence: \textit{t}(21) =  -2.19, \textit{p} $<$ .05) tasks. For the slingshot task, neither arousal nor valence were significantly different between the easy and challenging conditions (Arousal: \textit{t}(21) =  1.41, \textit{p}=0.174; Valence: \textit{t}(21) = 1.54, \textit{p}=0.159). However, one-tailed t-tests revealed that participants reported significantly higher valence on average for the slingshot task compared to the other three tasks (Button sequences: \textit{t}(86) =  3.20, \textit{p} $<$ .001; Card sequences: \textit{t}(86) =  3.91, \textit{p} $<$ .001;; Frustration: \textit{t}(86) = 4.37, \textit{p} $<$ .001;), suggesting that the task was comparatively successful at eliciting a sense of fun and enjoyment.


\begin{figure}[ht]
    \centering
    \includegraphics[width=0.7\linewidth]{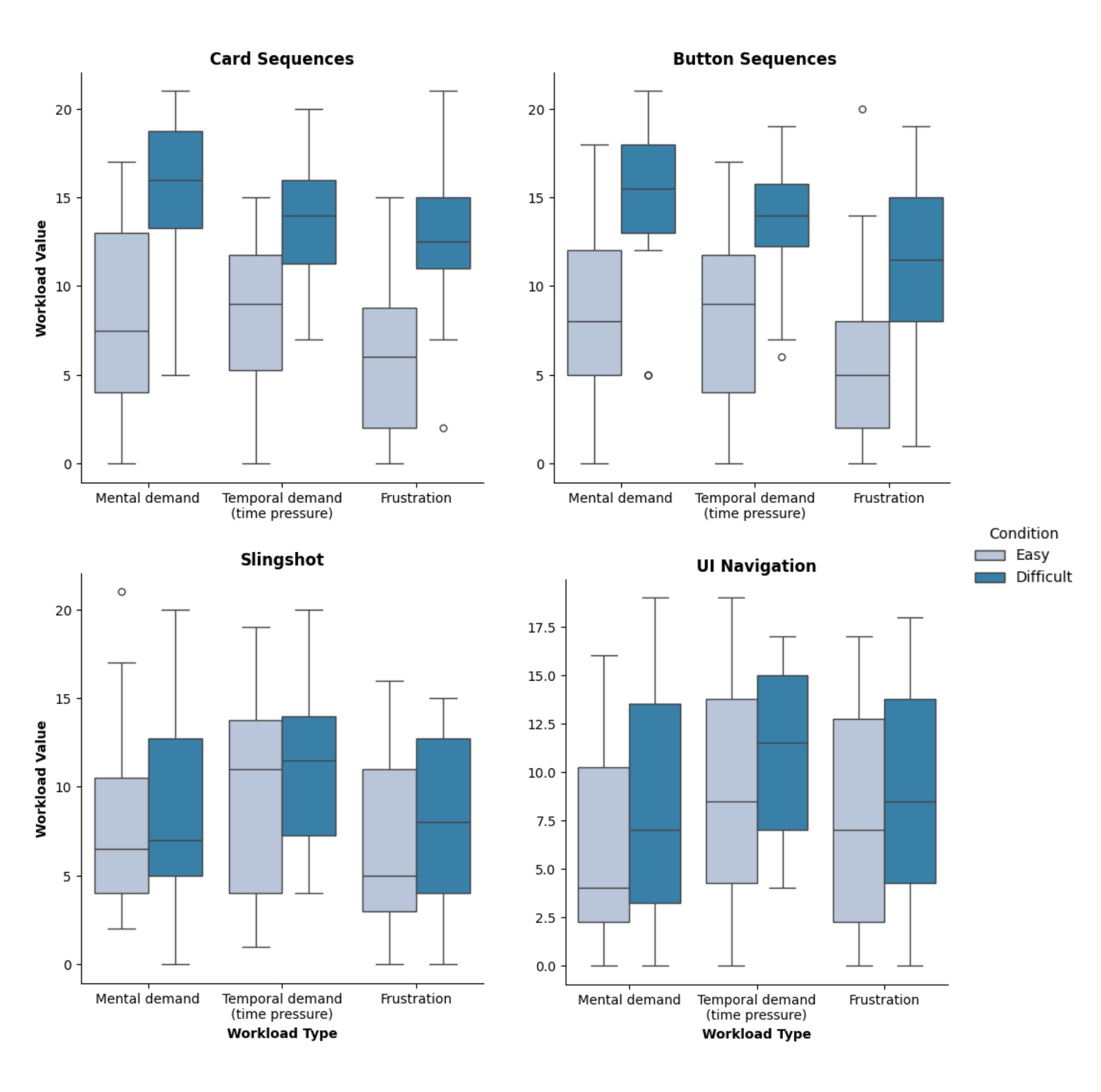}
    \caption{Effect of {\sc task condition} on induced mental workload in the \textit{easy} condition (gray) and \textit{challenging} condition (blue).}
    \label{fig:task-nasatlx}
    \Description{This image displays four box plots grouped by different tasks: "Card Sequences," "Button Sequences," "Slingshot," and "UI Navigation." Each plot compares workload values (y-axis) across three workload types (x-axis): "Mental demand," "Temporal demand (time pressure)," and "Frustration." Two conditions are represented: "Easy" (gray) and "Difficult" (dark blue). For all tasks, the "Difficult" condition tends to show higher workload values than the "Easy" condition across most workload types. "Mental demand" generally exhibits the highest workload values across tasks, particularly in the "Difficult" condition. As with the previous graph, the boxes show the interquartile range, with whiskers representing the range, and outliers indicated by small circles.}
\end{figure}

\textit{Mental workload with NASA-TLX. }We conducted one-tailed paired t-tests for the NASA-TLX measures across all four tasks. For the button sequence, card sequence and UI navigation tasks, mental workload (Card sequences: \textit{t}(21) =  5.63, \textit{p} $<$ .001; Button sequences: \textit{t}(21) =  4.18, \textit{p} $<$ .001; UI Navigation: \textit{t}(21) =  2.68, \textit{p} $<$ .01), temporal workload (Card sequences: \textit{t}(21) =  4.20, \textit{p} $<$ .001; Button sequences: \textit{t}(21) =  4.21, \textit{p} $<$ .001; UI Navigation: \textit{t}(21) =  1.92, \textit{p} $<$ .05) and frustration (Card sequences: \textit{t}(21) =  4.14, \textit{p} $<$ .001; Button sequences: \textit{t}(21) =  5.40, \textit{p} $<$ .001; UI Navigation: \textit{t}(21) =  1.78, \textit{p} $<$ .05) was significantly higher in the challenging condition compared to the easy condition. Although the direction of the results was the same for the slingshot task, none of the results reached significance (Mental workload: \textit{t}(21) =  .267, \textit{p} = .395; Temporal workload: \textit{t}(21) =  1.36, \textit{p} = .112; Frustration: \textit{t}(21) =  .675, \textit{p} = .276).




\subsubsection{Mental workload via Heart Rate Variability (HRV)}
We computed HRV indicators for each participant for each of the tasks and task conditions. A paired, two-tailed t-test was applied to the HRV indicators between the \textit{easy} and \textit{challenging} conditions. The results of the statistical tests are summarized in Table \ref{tab:hrv}. No HRV measure was found to be significantly different between the \textit{easy} and \textit{challenging} conditions for any task despite the differences in subjective self-report measures of affect and cognitive load described above (see supplementary materials for a further comparison of HRV measures between conditions and baseline). These results are consistent with previous findings that short-term HRV measures may not be sufficiently sensitive or robust to artifacts to reliably reflect affect and cognitive states \cite{sun2014moustress, potts2024sweating}.

\hrvtablenew

\subsection{Affect, Cognitive Workload and Gesture Formation} \label{sec:4.2}

\subsubsection{Gesture overview. } 
As a first step to analyzing the data, we aimed to describe the gesture formation process for each task. To this end, we draw on prior research that describes the primary phases of gesture formation:  \cite{ter2024hand, davis1994recognizing}: preparation (hands moving away from resting position), pre-stroke hold, stroke (the movement trajectories used to indicate commands) \cite{magrofuoco2021two}, post-stroke hold and retraction (hands moving towards resting position). Based on the recorded RGB video, we manually characterized movement during each gesture phase using a coding procedure adapted from prior research on nuances in gestures \cite{luo2024emotion}. In the initial phase, two researchers reviewed a sample of participant videos with a balanced distribution across tasks and conditions. The researchers independently employed a bottom-up approach to identify recurring patterns in gesture formation and noted down those deemed important or interesting. After both researchers had completed reviewing the sampled videos, they compared the patterns identified. Through several rounds of discussion, similar observations were merged and those that were less relevant or idiosyncratic were removed. The key observations for each task are summarized as follows.

\begin{figure}[ht]
    \centering
    \includegraphics[width=\linewidth]{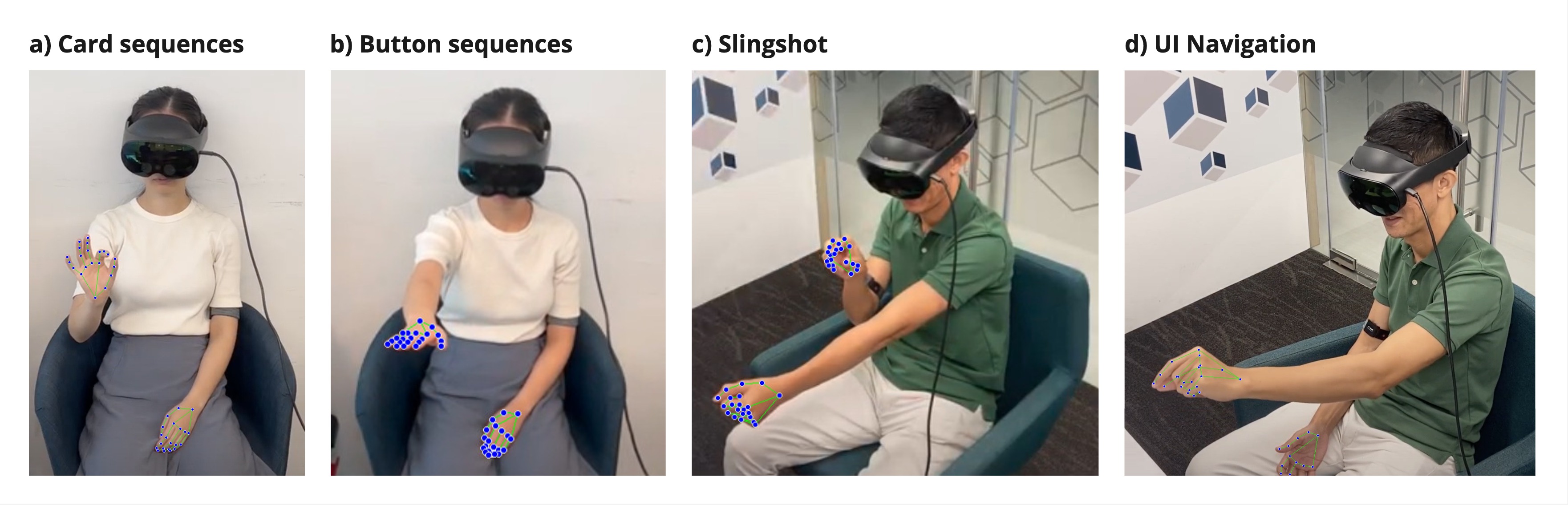}
    \caption{Illustration of detected hand keypoints during task performance. The RGB video and keypoints were used to manually characterize hand movements during the primary phases of gesture formation, including preparation (hands moving away from resting positions, typically on the lap), stroke (the movements used to indicate commands) and retraction (hands moving back towards resting positions).}
    \label{fig:hands-annotated}
    \Description{This image contains four side-by-side photos, each showing a participant wearing a VR headset and seated in a chair while performing tasks with hand gestures. Each photo is labeled with a task. a) Card sequences: The participant holds one hand up with fingers making a pinch gesture and the other hand resting on their lap. Blue dots and lines annotate the hand positions. (b) Button sequences: The same participant from (a) extends one hand forward, as if pressing virtual buttons, while the other hand rests on the lap. Hand gestures are also annotated with blue dots and lines. c) Slingshot: A different participant in a green shirt pulls one arm back as if using a virtual slingshot, with hand annotations highlighting their gesture. d) UI Navigation: the same participant as in (c) extends their arm, appearing to interact with a virtual interface in front of them. Both hands are annotated with blue dots and lines.}
\end{figure}

\begin{itemize}
    \item \textbf{Slingshot: } In the preparation phase, participants moved their hands from a resting position to pick up the slingshot.
    All but one participant held the slingshot in their non-dominant hand. The pre-stroke phase involved reaching forward with the dominant hand, pinching the index and thumb together to pick up a ball, and loading the ball into the slingshot. The stroke begins when participants pull their dominant hand back, as seen in Figure \ref{fig:hands-annotated}(b). Participants typically used their dominant hand to aim and angle, although use of the non-dominant hand was also observed for larger adjustments. The ball is shot by releasing the index finger and thumb apart. A short pause was observed after each shot, during which participants appeared to check whether or not they were successful at hitting the cups. Finally, in the post-stroke phase, the non-dominant hand is slightly retracted towards the body as participants prepare to reach for the next ball. No retraction was observed until the end of the task, and preparation only re-occurred if a participant dropped the slingshot.
    \item \textbf{Card sequences: } In the preparation phase, participants moved their dominant hand away from a resting position. For participants who performed selection with a pinching gesture, the pre-stroke phase consisted of moving the dominant hand in mid-air with the index and thumb slightly apart, allowing them to control a virtual pointer. A short pause was observed before the stroke, which appeared to allow participants to confirm the location of the virtual pointer while simultaneously stabilizing the selection gesture. The stroke (selection) involved bringing the index finger and thumb together in a pinch. In the post-stroke phase, the pinch is released and participants prepare to move to the next card. Retraction was observed at the end of each sequence; participants returned their dominant hand back to a resting position. 
    \item \textbf{Button sequences: } As with the previous tasks, preparation involved moving the dominant hand away from a resting position. The pre-stroke phase consisted of moving the hand in mid-air with the palm facing down to hover above a target button, as seen in Figure \ref{fig:hands-annotated}(a). Similar to the card sequences, we observed a short pause before the stroke was initiated that appeared to be participants confirming that they were correctly positioning their hand. The stroke involved bringing the hand down, with the fingertips slightly tilted downwards. Another short pause was observed after completion of the stroke. In the post-stroke phase, participants lift the hand back up and move towards the next target button. Retraction was observed at the end of each sequence.
    \item \textbf{UI navigation: } Following preparation, the pre-stroke hold involved holding the dominant hand in mid-air with the thumb and index finger slightly apart. The stroke (swipe) involved pinching the index finger and thumb together then flicking horizontally. In the post-stroke, participants released the pinch gesture and moved their dominant hand back to its original location in preparation for the next swipe. No retraction was observed until the end of the task. 
\end{itemize}

\begin{figure}[ht]
    \centering
    \includegraphics[width=\linewidth]{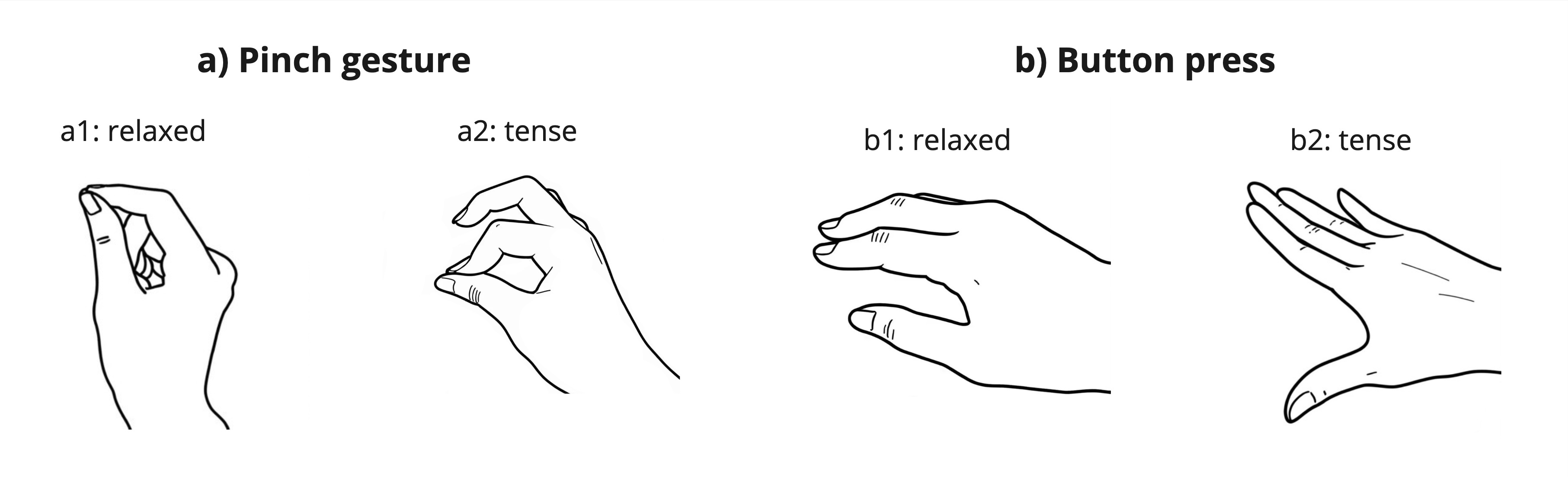}
    \caption{Illustrations of frequently observed versions of the same gesture: relaxed (left), and tense (right). Notably, gestures that look tense tend to be characterized by increased extension of the fingers.}
    \label{fig:gesture-tense}
    \Description{The image contains four sketches labeled a1, a2, b1 and b2. a1 and a2 show a hand making a pinch gesture, with a1 in a relaxed position and a2 showing a tense, flexed position. b1 and b2 show a hand making a motion as if to press a button. b1 similarly demonstrates a relaxed position, while b2 demonstrates a tense, flexed position.}
\end{figure}

Across several tasks, we observed the presence of implicit contextual information, defined by Schmidt \cite{schmidt2000implicit} as actions that are \textit{"...not primarily aimed to interact with a computerized system, but which such a system understands as input."} For example, the non-dominant hand was typically left in a resting position during the tasks that involved single-hand interactions, such as the button and card sequences (see Figure \ref{fig:hands-annotated}a, 5b). Even so, when participants were under high cognitive workload, we often observed that the resting hand gradually adopts a slightly flexed or tense pose (see Figure \ref{fig:hands-annotated}c). When the hand is relaxed, the fingers are generally in a neutral position with a slight inward curl in each knuckle and finger joint \cite{nielsen2003procedure}. As participants became more tense, we observed increased extension and hypterextension of the fingers with the specific joints involved depending on the hand gesture (Figure \ref{fig:gesture-tense}).

We also observed the use of expressive gestures unrelated to the task. For example, after receiving auditory feedback that an incorrect card was selected, P12 exclaimed: \textit{"Come on, I got that right!"}. At the same time, they gestured with both hands (palms open and supinated), in an expression of frustration and questioning.

\subsubsection{Gesture and motion features.}
Next, we explored methods to identify data segments containing meaningful gesture information. Across all tasks, the tip of the index finger on the dominant hand consistently recorded the most motion out of all the joints. Intuitively, all of the gestures used in this study primarily implicate the index finger. Previous work on free-hand gesture recognition has also focused on the trajectory of the index finger on the dominant hand \cite{caputo2018comparing} as a source of information. As such, in the following analyses, we primarily focus on the index finger tip to capture nuances in hand motion. Specifically, we analyze the wrist-relative motion of the index finger tip, which allows us to isolate gesture information from overall movements of the arm and body. The following analyses exclude the motion data of one participant (P4) due to technical issues during the recording.


From the hand tracking data, we extract gesture phases using the following procedure. The Savitzky-Golay digital filter was first applied to remove noise in the index tip movement signal (calculated as the Euclidean distance traveled over time). Then, a peak detection algorithm~\footnote{\url{https://docs.scipy.org/doc/scipy/reference/generated/scipy.signal.find_peaks.html}} was applied to the smoothed signal \cite{ye2015gestimator} to detect  periods of active movement. The height, width and prominence parameters of the algorithm were determined adaptively 
for each participant and task to account for idiosyncratic differences in movement patterns as well as the gestures required for each task. The video recordings were used as an additional visual guide to verify that relevant gesture phases could be detected. Using the left and right edges of each peak ( calculated as the left and right intersection points of a horizontal line at the peak's lowest contour line) to denote the start and end respectively of a gesture phase, we compute motion features such as speed and distance as well as pose features such as hand tension. We detected an average of 68.7, 48.6, 60.3 and 57.9 gesture phases per session for the card sequences, button sequences, UI navigation and slingshot tasks respectively. In the following paragraphs, we summarize our findings from the motion data. 

\textit{Gesture distance and speed. } Gesture distance, calculated as the total Euclidean distance traveled during the gesture phase, was significantly longer in the easy condition for the card sequence (\textit{t}(2184) =  6.70, \textit{p} $<$ .001) and button sequence (\textit{t}(1555) =  3.53, \textit{p} $<$ .001) tasks. Gesture speed was significantly faster in the easy condition for the card sequence (\textit{t}(2187) =  7.31, \textit{p } $<$ .001), button sequence (\textit{t}(1562) =  2.72, \textit{p} $<$ .01) and UI navigation (\textit{t}(1562) =  2.98, \textit{p} $<$ .01) tasks. 

\textit{Tension in the hand. } To analyze tension in the fingers and hands, we turned towards heuristics that were relevant for the set of gestures used in the study tasks. Heuristic-based approaches to gesture recognition and analysis are common and feasible, especially when the set of candidate gestures is small \cite{wobbrock2007gestures}. For the \textit{pinch} gesture used in the card sequences and UI navigation tasks, tense gesture exemplars typically featured increased pinch strength, leading to hyperextension of the distal interphalangeal joint of the index finger. As such, we calculated tension in the pinch gesture as the deviation of the distal interphalangeal joint from the straight line formed between the index finger tip and the proximal interphalangeal joint. A positive deviation (the distal interphalangeal joint lies above the line) was taken as an indicator of a more relaxed position, while a negative deviation was taken as an indicator of increased tension. We found that tension at the onset of each gesture phase was indeed higher in the challenging condition for both the card sequence (\textit{t}(2228) =  1.32, \textit{p} = .185) and UI nvaigation (\textit{t}(2451) = 1.30, \textit{p} = .194) tasks, though neither reached significance. For the \textit{button press} gesture used in the button sequences task, tense gesture exemplars typically featured increased extension of the entire finger, leading to hyperextension of the proximal interphalangeal joint of the fingers excluding the thumb. Similar to the pinch gesture, we calculated tension in the button press gesture as the deviation of the proximal interphalangeal joint from the straight line formed between the index finger tip and knuckle. Tension was significantly higher in the challenging condition (\textit{t}(1579) =  2.38, \textit{p} $<$ .05).

\textit{Head movement. } In addition to the hands, head movement also plays an important role in non-verbal communication and has been recognized as a valuable feature for distinguishing between different emotional expressions \cite{glowinski2011toward}. We calculated head movement as the total Euclidean distance moved during each gesture phase, and find that head movement was significantly reduced in the challenging condition for the card sequences (\textit{t}(2217) =  -4.10, \textit{p} $<$ .001), button sequences (\textit{t}(1562) =  -2.58, \textit{p} $<$ .05) and slingshot tasks (\textit{t}(2286) =  -2.17, \textit{p} $<$ .05). Although directionally similar, the difference in head movement for the UI navigation task did not reach statistical significance (\textit{t}(2328) =  -1.02, \textit{p} = .306).





\subsection{Classification Model}
\modeltab

To assess the baseline effectiveness of the extracted gesture and motion features described in section \ref{sec:4.2} for predicting affect and cognitive load, we employed support vector classification (SVC), a class of models that has been widely used for classification and regression tasks \cite{cervantes2020comprehensive}. The affect and cognitive load measures were simplified to two classes (low $\in$ [0,5], high $\in$ [6,10] for valence and arousal; low $\in$ [0,10], high $\in$ [11,21] for NASA-TLX measures of cognitive load) to remove noise and improve comparability with {\sc task condition}, which also had two classes (easy, challenging). We conducted an evaluation of prediction accuracy using two different approaches: (i) ten-fold within-user cross-validation, and (ii) leave-one-user-out cross validation. Cross-validation is a standard approach that assesses the predictive performance of models by using subsets of the data to iteratively train and test the model \cite{poldrack2020establishment}. Using the leave-one-user-out approach, we also evaluate the ability of the models to generalize to novel exemplars. For within-user evaluations, we trained the model of 90\% of each participants' data and tested it on the remaining 10\% of the data, following the approach outlined in \cite{wang2024watch}. We used the scikit-learn implementation of SVC \cite{chan2022reporting} and tested multiple regularization terms ($10^{-1}$, 1, 10, 20). We evaluate the models in terms of accuracy and F1 score. Model performance for each approach is summarized in Table 2 (see supplementary materials for additional metrics).

The best-performing models for within-user cross-validation achieved an average accuracy of 88\% for the card sequences task, 85\% for the button sequences task, 91\% for the UI navigation task and 87\% for the slingshot task across the cross-validation folds when predicting overall task condition. F1-scores indicate that the models achieve good precision and recall ($>$ 0.80). Recognition of affect and cognitive load was more varied, ranging from 50\% (chance accuracy) to 87\%. With leave-one-user-out cross-validation, model performance generally decreases, suggesting that simple SVCs struggle to generalize to targets it has not seen before. Considering the percentage of the most frequently occurring class as baseline as well as the F1 scores, we also find that SVCs occasionally face difficulty with predicting the minority class. This issue is especially apparent with recognition of the mental demand dimension of the UI navigation and Slingshot tasks, possibly due to the similar distribution of mental demand between the easy and challenging conditions for these two tasks (see Figure \ref{fig:task-nasatlx}). As a first step, our baseline models show that gesture and motion features can be predictive of affect and cognitive load, but capturing complex or subtle relationships may require more sophisticated feature extraction and modelling methods.

\section{Discussion}
We studied the potential of using free-hand gesture interactions for inferring affect and cognitive load in virtual reality. The literature on inferring mental states from user inputs emphasizes the visceral relationship between the mind and the body, especially as it relates to movement. Therefore, our work extend prior findings in this area to free-hand gesture inputs in VR, which utilizes very different interaction techniques and sensor inputs compared to conventional computing devices such as PCs, laptops and mobile devices. In this section, we reflect on the connection between mental states such as affect and cognitive load, and subtle changes in the pattern of hand and head motion during gesture formation. We then discuss the implications of these findings for designing gesture-based user input systems.


\subsection{Gesture distance and speed } Our exploration of gesture phases showed that gesture distance and speed serve as statistically significant indicators of participants being in the challenging task condition. Specifically, we find that participants tend to form gestures at a slower speeds when under challenging task conditions. These results align with prior work on inferring mental states from user inputs as well as the predictions of Fitts' law and the Hick-Hyman law \cite{hibbeln2017your, grimes2015mind, fitts1954information, hick1952rate, hyman1953stimulus, burno2015applying}. 

In terms of distance, however, our results were surprising. We find that participants tended to reduce the size of the movements required to form a gesture (i.e., shorter gesture distances) when under challenging task conditions. This finding was robust to different peak-detection algorithm parameters and removal of the preparation and retraction phases, and held regardless of whether absolute or wrist-relative motion was considered (see supplementary materials for detailed results). Based on prior work, we would expect the opposite: attentional control theory (ACT) \cite{eysenck2007anxiety} predicts that the lower valence induced by the challenging conditions should result in reduced attentional control, thereby increasing deviations from one's intended movement trajectory and the overall distance traveled during gesture formation. Similarly, the Hick-Hyman law predicts that the high cognitive load induced by the challenging conditions should cause increased variability in fine motor control, hence increasing the overall distance traveled during gesture formation.

To understand why theoretical predictions for gesture distance failed to hold for free-hand gesture interactions in VR, we consider the question: what makes free-hand interactions different from interactions that rely on tangible input tools, such as computer mice? One possible explanation comes from prior research which demonstrates that cognitive load and mental stress increase muscle stiffness \cite{sun2014moustress}. With traditional input devices such as the mouse, studies are designed such that participants must manipulate the mouse to move the cursor a certain distance across the screen, with the minimum distance the participant must move determined by the dots per inch (DPI) of the mouse. In this context, increased muscle stiffness translates to increased mouse movement due to decreased precision when moving the mouse. However, with gesture inputs, users are manipulating their hands instead of an external tool and are free to form their gestures in various ways as long as the command gesture is eventually recognized. For example, with the pinching gesture used for selection, participants could hold their thumb and forefinger very close together between selections for greater efficiency, or could "bounce" the two fingers off each other, generating more movement. As such, increased muscle stiffness in the context of free-hand gestures could potentially correspond to reduced extension and flexion \cite{van2001changes} of the joints implicated in gesture formation, leading to shorter gesture distances overall. The hypothesis that participants experience greater muscle stiffness when experiencing higher cognitive load is also supported by our findings regarding increased tension in the hand in the challenging condition over several tasks. 

\subsection {Head motion } In addition to gesture formation, we also explored head movements captured from the VR headset as an avenue to recognise affect and cognitive load. Although prior work has examined the use of head movements to \textit{communicate} emotional expressions, whether and why factors such as affect and cognitive load might influence head motion in a manner below the level of conscious awareness remains unclear. In our study, we found that the challenging conditions were associated with significantly reduced head movement for all but one of the tasks. We turn again to theories of attention and the Hick-Hyman law to reflect on this finding. It is generally accepted that the brain receives more input than it can process, and that attention selects which stimuli or actions get access to these capacity-limited processes \cite{teasdale2001attentional}. Furthermore, processing of proprioceptive information (our sense of the positions of our limbs relative to the body) and vestibular information (our sense of rotational movement and linear acceleration) has been shown to consume cognitive resources \cite{ bigelow2015vestibular}. We speculate that when experiencing high cognitive load, it is not only the bandwidth for fine motor control that is impacted. Instead, to further increase the mental capacity available for the task at hand, the brain also instinctively reduces incoming information from the proprioceptive and vestibular systems by holding still. This interpretation of the results also helps to explain why gesture distance was found to be reduced in the challenging condition -- that in addition to increased muscle stiffness, participants may also have subsonsciously tried to reserve mental capacity for working on the task by decreasing the size of their movements.

\subsection{Comparing motion vs. physiological data }
Our findings highlight the potential of hand and head motion as an alternative objective measure of affect and cognitive load in tasks that involve free-hand gesture inputs. Physiological signals such as heart rate variability (HRV) are widely used as indicators of affect and cognitive load, as well as other metrics of interest such as stress and sleep quality \cite{hall2004acute}. Despite significant differences in self-reported ratings of affect and cognitive load between different task conditions in our study, no HRV indicator was found to be significantly different when comparing the \textit{challenging} condition to the \textit{easy} and \textit{baseline} conditions, echoing findings from previous work that used short-term or ultra-short-term (UST) HRV measures \cite{sun2014moustress,potts2024sweating}. Short-term HRV measures have become an increasingly attractive method of measuring participant affect and cognitive load in recent years due to the emergence of affordable wearable sensors in convenient form factors, and the promise of more objective data compared to subjective self-reports. Yet, the physiological processes that generate changes in short-term HRV are still not well-understood, compared to longer recordings where change is known to be driven by processes such as circadian rhythm and fluctuations in core body temperature \cite{shaffer2020critical}. HRV measures also tend to be noisy and susceptible to motion artifacts, making it challenging to collect high-quality data when research tasks require participants to move around. 

%

\subsection{Design Implications and Future Work}
Our work opens up opportunities for extended reality systems that utilize gesture-based user inputs to support recognition of affect and cognitive load. In the following, we discuss two application scenarios that could benefit from affect and cognitive load predictions from gesture inputs.

\textit{Adaptive interfaces for productivity support. } Augmented Reality (AR) devices are increasingly being used to support productivity tasks. For example, AR is often used during complex assembly processes due to its ability to overlay digital instructions onto objects in the real world, reducing the split-attention effect that otherwise occurs when individuals repeatedly switch attention between assembly instructions and the physical task \cite{funk2015benchmark, wang2022comprehensive}. AR has also been investigated in the context of complex learning tasks such as surgical procedures \cite{andersen2016effect}. Gesture inputs can be used to develop adaptive systems that dynamically modify the delivery of task content based on the users' affect and cognitive load, an approach which has been shown to effectively encourage learning, support learning achievement and reduce task anxiety \cite{hwang2020fuzzy} compared to conventional systems. 

\textit{Augmented communication. } Technology-mediated communication interfaces have become ubiquitous, allowing users to connect with others remotely. However, digital communications are often challenging due to reduced access to important nonverbal and expressive cues. Social virtual reality spaces such as AltspaceVR and VRchat have begun to address these limitations to an extent, by providing immersive experiences that support both verbal and nonverbal communication \cite{maloney2020falling}. Another common theme that threads through efforts to augment the digital communication experience is the use of biosignals: Significant Otter \cite{liu2021significant} investigated the use of animated otter avatars for romantic couples to share and respond to each others' biosignals, and found that it enhanced participants' ability to communicate and connect with their partner. Similarly, HeartChat \cite{hassib2017heartchat} presented a mobile application with heart rate augmented chats, and found that sharing physiological information supported empathy and awareness of the chat partner's context and emotional state. Our work demonstrates that, like biosignals, patterns of user hand and head motion also provide information about affect and cognitive load and may potentially be able to capture more subtle changes than commonly-used physiological signals. Additionally, our approach does not require any additional wearable sensors or modifications to a standard VR headset. Motion-based indicators of affect might hence be an accessible way to augment communication experiences in social VR. 

Although automatic recognition of affect and cognitive load holds promise for the development of systems that can respond more intelligently to the user, we are also aware of the personal and private nature of  mental states and the ways in which such technologies could be misused. For example, in a recent review of mental health applications, \citet{kang2024app} discuss how lack of guidance on how to interpret information about one's mental states can lead to feelings of fear and distress. A mismatch between detected states and how the user subjectively feels could also influence their personal judgement, especially when users believe that a system has access to privileged information about them \cite{hollis2018being}. As such, when using sensing technologies, we believe that designers should carefully consider how the outputs of such technologies are presented, keeping the target audience in mind.









\section{Limitations \& Future Work}
While our findings demonstrate the potential of using gesture, hand and head motion features to infer cognitive load and affective states in virtual reality, there are several limitations to address in future work.

We explored relatively simple methods of extracting features from the hand and head tracking data that partially relied on human judgement. Additionally, in this study, we primarily analyzed the index finger as a source of gesture and hand motion information as it was responsible for the majority of the movement in all our tasks. To improve the input features for modeling and expand on our insights, future work can consider using more sophisticated gesture recognition and segmentation methods to extract a wider range of gesture and motion features, as well as explore the role of other fingers and joints in the hands and arms. Specialized hand tracking systems such as the Leap Motion controllers \footnote{\url{https://www.ultraleap.com/}} can also be used to validate our results.

Although our selected tasks and task conditions successfully induced the intended affect and cognitive load, they also tended to alter multiple factors simultaneously. For example, the challenging conditions often increased not just mental demand, but also temporal demand and frustration, making it difficult for quantitative methods to identify the specific dimensions of affect or cognitive load that were contributing to observed differences in the features extracted. Future work can address this by developing more refined task designs that isolate different types of affective and cognitive effects, allowing for a more detailed understanding of how specific dimensions influence gesture, hand and head motion. 

Although our basic classification model suggests that gesture and motion features are indicative of affect and cognitive load, as yet, the models may not perform well enough to be used in real-world systems. Our goal in the current work was to show that affect and cognitive load induce significant differences in the pattern of users' hand and head motion as they interact with VR environments. Given the strong correlation, even a basic model with minimal  parameter optimization is able to classify affect and cognitive load with reasonable accuracy. Going forward, we plan to expand this work by augmenting our dataset with more tasks and participants, and use more sophisticated machine learning models that can learn complex relationships in the data to improve classification results.

\section{Conclusion}
We present \textit{Motion as Emotion}, a novel method to recognise affect and cognitive load using hand and head motion as the sole source of information. We conducted a study to compare gesture distance, speed, hand tension and head motion as participants complete tasks in conditions that induce different levels of valence, arousal and cognitive load. Results showed that patterns of hand and head movement change significantly across conditions, and that standard support vector machines could classify task condition with up to 91\% accuracy with minimal fine-tuning. We provide detailed insights on how gestures are formed in the context of free-hand gesture inputs in virtual reality environments, explain how cognitive and affective states might influence the gesture formation process through the mechanisms of muscle stiffness and attentional control. We discuss the implications of enabling recognition of affect and cognitive load through free-hand gestures in different scenarios, and make recommendations for future research and design.


\begin{acks}
\end{acks}

\bibliographystyle{ACM-Reference-Format}
\bibliography{references}


\end{document}